\documentclass[conference]{IEEEtran}
\IEEEoverridecommandlockouts
\usepackage{cite}
\usepackage{amsmath,amssymb,amsfonts}
\usepackage{graphicx}
\usepackage{textcomp}
\usepackage{xcolor}
\usepackage{algorithm}
\usepackage[caption=false,font=footnotesize,captionskip=-2pt]{subfig}

\usepackage{algpseudocode} 
\usepackage{booktabs}
\usepackage{bm}

\IEEEsettextwidth{0.625in}{0.625in}
\IEEEsetsidemargin{c}{0pt}

\def\BibTeX{{\rm B\kern-.05em{\sc i\kern-.025em b}\kern-.08em
    T\kern-.1667em\lower.7ex\hbox{E}\kern-.125emX}}

\begin{document}

\title{Backscatter Assisted Indoor NLOS Positioning
}

\author{
    \IEEEauthorblockN{Kalle Ruttik, Hüseyin Yiğitler, Jingyi Liao, Alexander Sheverdyaev, and Riku Jäntti}
    \IEEEauthorblockA{
        Department of Information and Communications Engineering\\ Aalto University\\ Espoo, Finland\\ }
    \thanks{This work was supported in part by the EU SNS project Ambient6G under Grant 101192113 and in pert by Research Council of Finland project BANLOS under Grant 375519. }
}

\maketitle

\begin{abstract}
Passive backscatter devices (BDs) can enable indoor non-line-of-sight (NLOS) positioning by serving as virtual anchors
whose Doppler-separated signatures are observable in standard channel estimates. This paper studies continuous
user-equipment (UE) tracking in corridor environments using a noncoherent power-domain formulation that avoids BD phase
synchronization and remains robust to residual carrier offsets and strong multipath. The BD-dependent measurements are
modeled by a log-distance law with unknown BD-specific offsets, which allows passive asynchronous devices to be used as
anchors without transmit-power calibration. Based on this model, we develop a corridor-constrained maximum a posteriori
(MAP) tracker with motion regularization and Huber-robust estimation. In ray-tracing-inspired simulations, the method
achieves median positioning errors of 0.23--0.27~m with 90th-percentile errors below 0.45~m. In office-corridor
measurements with four passive BDs at 866~MHz, it attains an aggregated median error of 0.505~m and outperforms a
simple weighted-average baseline. The results show that passive asynchronous BDs can provide practical
sub-meter indoor NLOS tracking while remaining compatible with existing channel-estimation pipelines and
energy-autonomous BD deployments.
\end{abstract}


\section{Introduction}
Positioning and localization are becoming increasingly important capabilities of modern mobile networks, enabling
emergency services, industrial automation, asset tracking, and digital-twin applications~\cite{abuyaghi2025positioning}.
While current cellular systems can achieve sub-meter accuracy under favorable line-of-sight (LOS) conditions, robust
performance in non-line-of-sight (NLOS) environments, including indoor corridors, urban canyons, and tunnels, remains
challenging. In such settings, blockage, diffraction, and multipath often limit the usefulness of conventional time- and
angle-based measurements, especially when bandwidth and antenna aperture are constrained.

Cellular positioning typically relies on standardized reference signals, such as positioning reference signal (PRS) and
sounding reference signal (SRS), to extract timing, angle, and power measurements~\cite{yang2022overview}. Although
these can provide high accuracy in LOS conditions, NLOS mitigation remains a major bottleneck in 5G and beyond
positioning, motivating work on multipath-assisted methods, fingerprinting, cooperative positioning, and programmable
propagation control~\cite{abuyaghi2025positioning}. However, many such approaches require wide bandwidths, large antenna
arrays, dense calibration data, or substantial infrastructure coordination, which can limit scalability in practical
indoor deployments.

This paper investigates \emph{backscatter-assisted positioning} as a complementary and infrastructure-light approach to
indoor NLOS tracking. The key idea is to deploy passive backscatter devices (BDs) at known locations so that they act as
engineered \emph{virtual anchors}. By modulating and reflecting an external illumination signal, these devices create
location-dependent signatures that can be observed in standard receiver channel estimates, without active transmission
from the anchors themselves. The studied scenario is compatible with the bi-static connectivity topology considered for
ambient Internet of Things (A-IoT) systems \cite{al2025ambient}, where an external transmitter provides the illumination signal and passive devices
interact with the network through backscatter. Moreover, the proposed approach is applicable to existing Long-Term Evolution (LTE) and New Radio (NR)
systems by exploiting standard reference signals instead of a dedicated continuous-wave carrier. In such systems,
receiver channel estimation can extract the backscatter components, thereby enabling detection and processing without
radio hardware modifications, as discussed in \cite{jantti2025integration}.

Prior work on backscatter localization has mainly focused on localizing the tags themselves, often using dedicated
readers, calibrated infrastructure, or phase-coherent
processing~\cite{soltanaghaei2021tagfi,xu2023principle,bae2024supersight}. In contrast, using passive BDs as anchors at known locations for positioning a separate user device remains largely unexplored. The closest
prior concepts either rely on synchronization and calibration~\cite{nokia2022backscatter,qualcomm2024backscatter} or
provide only coarse position estimates based on the identity of detected BDs~\cite{elsanhoury2025zero}.

\addtolength{\topskip}{3pt}
Specifically, we consider an indoor corridor NLOS scenario with a single illuminating transmitter, multiple passive BDs
at known locations, and a receiver operating under noncoherent processing constraints. Within this setting, we formulate
a noncoherent dB-domain measurement model with unknown BD-dependent offsets and a common path-loss exponent, and develop
a robust corridor-constrained maximum a posteriori (MAP) /Viterbi-tracker that combines motion regularization with Huber-based estimation. The
approach is evaluated in both simulation and measurement: the simulation study uses a ray-tracing-inspired L-shaped
corridor model together with a smoothed ray-tracing benchmark to quantify channel-model mismatch, while the measurement
study uses an office-corridor campaign. 
The framework is instantiated using a simple single-slope log-distance fingerprint model, but it is not restricted 
to this choice;
richer environment-aware fingerprints, for example derived from a digital twin, could be incorporated within the same
tracking architecture. To the best of our knowledge, this work provides one of the first experimental demonstrations of
continuous user equipment (UE) positioning using passive asynchronous BDs deployed
as known-location virtual anchors.

The rest of the paper is organized as follows. Section II introduces the system model and measurement formulation.
Section III presents the proposed tracking algorithm. Section IV reports simulation results, and Section V presents
measurement results. Section VI discusses the relevance of the considered scenario to A-IoT
deployments and the applicability of the proposed approach to LTE and NR systems. Section VII concludes the paper.

\section{System model}
We consider a \emph{bi-static backscatter} positioning setup in an indoor corridor, where a set of $K$ BDs with known
locations $\{\bm{B}_k\}_{k=1}^K\subset\mathbb{R}^2$ act as \emph{virtual anchors}. An indoor transmitter emits a continuous carrier wave
(CW) that illuminates the corridor environment. Each BD modulates the incident carrier by applying a small frequency
shift, which enables the receiver to separate BD-dependent backscatter components from the direct carrier component in
the frequency (Doppler) domain using standard channel-estimation processing.

We consider a NLOS operating condition in which no direct propagation path exists between the transmitter and the
receiver. The transmitter--BD links are also generally NLOS due to indoor blockage. Because backscatter propagation
experiences the combined attenuation of the transmitter--BD and BD--receiver links, the detectable range of passive BDs
is typically limited. In practical deployments, BDs are therefore placed in close proximity to the tracking area so that
their reflections remain observable by the receiver. Accordingly, we consider a localized corridor region where the
receiver can detect all BDs over the observation window. Such a configuration is representative of indoor deployments in
which a small number of nearby passive devices augment positioning performance within a bounded area.

In practical scenarios, the BDs are not synchronized and do not share a common phase reference. In addition, residual
carrier-frequency offsets typically exist between the transmitter and the receiver due to imperfect synchronization.
Combined with NLOS multipath propagation, these effects make coherent phase-based localization unreliable. We therefore
adopt a \emph{noncoherent} formulation based on power-domain measurements rather than phase-coherent signal models.

Let $\bm{U}(t)\in\mathbb{R}^2$ denote the UE position at time $t$. The received BD-dependent power arises from a bistatic propagation
chain consisting of a transmitter--BD link and a BD--UE link. Since the transmitter and BD locations are fixed, the gain
of the transmitter--BD link is constant over time and can be absorbed into a BD-specific offset parameter. Consequently,
the temporal variation of the received power is dominated by the BD--UE link, which depends on the UE position. We
therefore approximate the position-dependent component of the received power using a single-slope log-distance
surrogate~\cite{rappaport2002wireless}
\begin{equation}\label{eq:measurement-k}
Z_k(t) = G_k - 10 p \log_{10}\!\bigl(\|\bm{B}_k-\bm{U}(t)\|_2\bigr),
\end{equation}
\noindent where $p$ represents the path-loss exponent of the BD--UE link and is assumed to be common across
BDs. The parameter $G_k$ is a BD-specific unknown offset that is constant over $t$. It absorbs all stationary link
components, including the fixed transmit power, antenna gains, the transmitter--BD propagation loss, and other
link-invariant effects. Since $G_k$ is device-dependent, each BD can be viewed as an \emph{effective transmitter with
unknown power}. This model captures the large-scale distance-dependent trend of the received power,
while residual deviations caused by multipath fading are included in the next section.

Estimating $\bm{U}(t)$ from the measurements $\{Z_k(t)\}$ is therefore closely related to received-signal-strength (RSS)
localization with unknown transmit power~\cite{huang2016rss}. Because a single measurement snapshot contains $K$
observations but $K+3$ unknown parameters ($\{G_k\}$, $p$, and the two-dimensional position), the instantaneous problem
is under-determined. To resolve this, we utilize a window of $T$ consecutive measurements. By assuming the offsets
$\{G_k\}$ and the exponent $p$ are quasi-static, and by applying local-motion regularization to the UE trajectory, the
accumulated observations provide sufficient degrees of freedom for joint estimation.

\section{Tracking algorithm}
\label{sec:tracking}

We estimate the UE trajectory $\{\bm{U}(t)\}_{t=1}^T$ from the frequency-separated BD power measurements
$\{Z_k(t)\}_{k=1}^K$ by formulating tracking as a discrete-state MAP sequence estimation problem on
a \emph{corridor-constrained grid}.

We start by discretizing the 2D region of interest into square cells (pixels) of side length $\Delta$ with centers
$\{\bm{U}_\ell\}_{\ell=1}^{L}\subset\mathbb{R}^2$. The UE position at time $t$ is approximated by the center of the
corresponding pixel, i.e., $\bm{U}(t)\approx \bm{U}_{\ell_t}$, where $\ell_t$ denotes the pixel index. This
discretization allows us to define admissible locations: cells outside the corridor or inside walls are marked using a
binary valid-state indicator $v_\ell\in\{0,1\}$, where $v_\ell=0$ denotes an inadmissible cell and $v_\ell=1$ an
admissible one.

The log-distance model becomes ill-defined when the UE coincides with a BD location. To avoid this singularity, path
lengths are regularized by adding a small constant $d_0>0$. We define the dB-domain fingerprint as
\begin{equation}
\label{eq:log-distance-fingerprint}
\Delta_{\ell,k}(p)\triangleq -10 p \log_{10}\!\big(\|\bm{U}_\ell-\bm{B}_k\|_2 + d_0\big).
\end{equation}

The cost function for the MAP sequence estimator can be written as the negative log-posterior of the measurements:
\begin{equation}
J \propto -\ln \mathrm{P}\big(\{Z_k(t)\}_{k,t} \mid \ell_{1:T}\big) - \ln \mathrm{P}(\ell_{1:T}),
\end{equation}
where $\mathrm{P}(\cdot)$ denotes the probability density or mass function. Under the measurement model 
\begin{equation}\label{eq:measurement-model-huber}
Z_k(t)\approx G_k+\Delta_{\ell,k}(p)+\varepsilon_k(t),
\end{equation}
the UE is in pixel $\ell$ at time instant $t$, $G_k$ is an unknown BD-dependent offset (in dB) absorbing link-invariant
gains/losses, and $\varepsilon_k(t)$ captures residual deviations. Since the measurement noise may deviate from a strict
Gaussian distribution, we adopt the Huber loss with threshold $\delta$, which corresponds to the least-favorable
distribution in Huber's robust estimation framework.

\begin{algorithm}[b!]
\footnotesize
\begin{algorithmic}[1]
\Require Power measurements $Z_k(t)$ for $t=1,\dots,T$, $k=1,\dots,K$
\Require BD locations $\bm{B}_k$ and grid cell centers $\bm{U}_\ell$, $\ell=1,\dots,L$
\Require Valid-state indicator $v_\ell$ and neighbor sets $\mathcal{N}(\ell)$ (radius $R_{\mathrm{move}}$)
\Require Candidate set $\mathcal{P}$ for $p$, Huber thresholds $\delta$ and $\delta_{\mathrm{move}}$
\Require Motion weight $\lambda_{\mathrm{move}}$, outer iterations $N_{\mathrm{out}}$, IRLS iterations $N_{\mathrm{IRLS}}$
  \State Precompute the geometric term used in \eqref{eq:log-distance-fingerprint}
\State $J_{\min}\gets +\infty$
\For{each $p\in\mathcal{P}$}
    \State Form $\Delta_{\ell,k}(p)$ from \eqref{eq:log-distance-fingerprint}
  \State Initialize $G_k^{(p)} \gets \mathrm{median}_t\, Z_k(t)$ for $k=1,\dots,K$
  \For{$i=1,\dots,N_{\mathrm{out}}$}
    \State Set $c_t(\ell)\gets +\infty$ for all $\ell$ with $v_\ell=0$
    \State Compute emission costs using \eqref{eq:emission-cost}
    \State Run Viterbi/DP over $\mathcal{N}(\ell)$ using \eqref{eq:motion-cost} to obtain $\ell_{1:T}^{(p)}$
    \For{$k=1,\dots,K$}
    \State Update $G_k^{(p)}$ by $N_{\mathrm{IRLS}}$ Huber-IRLS steps under \eqref{eq:measurement-model-huber}
    \EndFor
  \EndFor
  \State Evaluate $J \gets J(p,\bm{G}^{(p)},\ell_{1:T}^{(p)})$ using \eqref{eq:objective_grid}
  \If{$J<J_{\min}$}
    \State Store best $(\hat p,\hat{\bm{G}},\hat\ell_{1:T})\gets(p,\bm{G}^{(p)},\ell_{1:T}^{(p)})$
    \State $J_{\min}\gets J$
  \EndIf
\EndFor
\State \Return $\hat{\bm{U}}(t)=\bm{U}_{\hat\ell_t}$, $\hat p$, and $\hat{\bm{G}}$
\end{algorithmic}
\vspace{3pt}
\caption{Robust corridor-constrained grid tracking with offset profiling and pathloss search}
\label{alg:gridtracker}
\end{algorithm}

For this measurement model, minimizing the negative log-posterior yields the robust cost function
\begin{equation}
\label{eq:objective_grid}
J(p,\bm{G},\ell_{1:T}) = \sum_{t=1}^{T}c_t(\ell;p,\bm{G}) 
 + \sum_{t=2}^{T} m(\ell_{t-1},\ell_t),
\end{equation}
where the emission cost at time $t$ for cell $\ell$ is defined as
\begin{equation}
c_t(\ell;p,\bm{G}) \triangleq
\sum_{k=1}^{K}\rho_\delta\!\left(Z_k(t)-G_k-\Delta_{\ell,k}(p)\right),
\label{eq:emission-cost}
\end{equation}
with $c_t(\ell;\cdot)=+\infty$ if a pixel is inadmissible ($v_\ell=0$), and 
$\rho_\delta(\cdot)$ is the Huber loss with threshold $\delta>0$, and
$\bm{G} \triangleq [G_1, \cdots, G_K]^\top$. 

The penalty term $m(\ell',\ell)$ in \eqref{eq:objective_grid} is proportional to the negative log-transition probability
$-\ln \mathrm{P}(\ell_t \mid \ell_{t-1})$, and increases with the distance
$\|\bm{U}_\ell-\bm{U}_{\ell'}\|_2$. Since we estimate the complete trajectory, this transition is
Markovian. We impose this first-order Markov structure through pairwise transitions
$\mathrm{P}(\ell_t \mid \ell_{t-1})$ together with admissibility constraints between consecutive time indices.
Although more involved transition costs can be defined, in this work, it is defined as
\begin{equation}
m(\ell',\ell)=
\begin{cases}
\lambda_{\mathrm{move}}\,\rho_{\delta_{\mathrm{move}}}\!\left(\|\bm{U}_\ell-\bm{U}_{\ell'}\|_2\right),
& \text{admissible}\\
\infty, & \text{otherwise}
\end{cases}
\label{eq:motion-cost}
\end{equation}
where $\lambda_{\mathrm{move}}>0$ is the motion-regularization weight and $\delta_{\mathrm{move}}>0$ is the Huber threshold 
used in the motion term. A transition $(\ell',\ell)$ is admissible if both cells are valid, $v_{\ell'}=v_\ell=1$, and if
$\|\bm{U}_\ell-\bm{U}_{\ell'}\|_2 \le R_{\mathrm{move}}$, where $R_{\mathrm{move}}>0$. Thus, $R_{\mathrm{move}}$ defines the local
neighborhood $\mathcal{N}(\ell)$ and allows transitions only between admissible cells within radius $R_{\mathrm{move}}$
(including self-transitions). In the numerical studies, we set $\delta_{\mathrm{move}}=\Delta$. 

For fixed $\{p,\bm{G}\}$, the objective is to estimate the discrete state sequence $\ell_{1:T}$ that minimizes
\eqref{eq:objective_grid}. Since the cost function involves discrete states and Markovian transitions, the minimizer is
obtained via Viterbi-style dynamic programming over $\mathcal{N}(\ell)$~\cite{trogh2015advanced}.

The remaining task is to estimate $p$ and $\bm{G}$. While an expectation-maximization (EM) approach is possible, we use
a simpler alternating procedure with Huber iteratively reweighted least squares (IRLS) updates for the offsets. To avoid over-tuning and accommodate
environmental variability, $p$ is selected by a coarse grid search over a candidate set $\mathcal{P}$. For each
$p\in\mathcal{P}$, the offsets $\bm{G}^{(p)}$ are initialized by the per-BD temporal medians of the observations. We
then compute $\Delta(p)$, run dynamic programming to obtain $\ell_{1:T}^{(p)}$, update $\bm{G}^{(p)}$ using
$N_{\mathrm{IRLS}}$ IRLS steps, and re-evaluate the objective. This loop repeats for at most
$N_{\mathrm{out}}$ outer iterations, with early stopping when the relative objective change becomes negligible. Finally,
the candidate $p$ with the lowest final objective is retained. The complete procedure is summarized in
Algorithm~\ref{alg:gridtracker}.

\section{Simulation results}
\label{subsec:rt_model}

\begin{figure}[t]
\centering
\includegraphics[width=0.8\columnwidth]{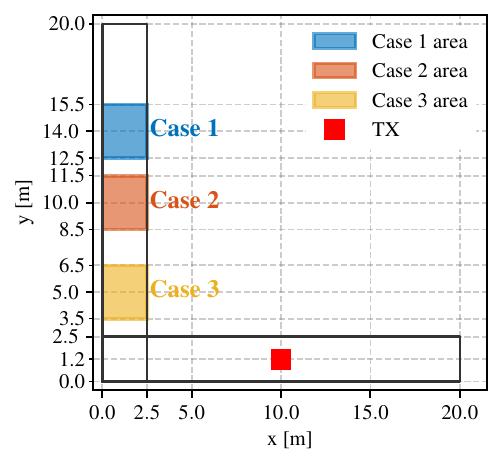}
\vspace{-5pt}
\caption{L-shaped corridor layout and the study case regions}
\label{fig:layout}
\end{figure}

To evaluate the proposed tracker under realistic NLOS power fluctuations, we generate synthetic measurements using a
ray-tracing-inspired wideband channel model in a 2-D L-shaped corridor, shown in Fig.~\ref{fig:layout}. The layout
contains a fixed transmitter (TX) illuminating the corridor and a set of wall-mounted BDs with
known locations. The UE moves along the vertical corridor leg, while NLOS propagation between the corridor legs is
mediated by a dominant corner diffraction region.

We study three areas (Cases~1--3) to represent distinct operating conditions. In Cases~1 and~2, the BDs are deployed
sufficiently far from the diffraction edge, so that the incident (illumination) power varies only mildly along the
corridor wall and the BDs experience relatively uniform illumination. In Case~3, the BDs are placed closer to the
diffraction edge, which leads to significantly stronger spatial variations in illumination power.
Cases~1 and~2 further differ in their overall illumination level: in Case~2 the BDs are, on average, farther 
from the transmitter than in Case~1, resulting in a lower illumination power.

For each propagation link between two points (TX$\rightarrow$UE, TX$\rightarrow$BD, and BD$\rightarrow$UE), the
ray-tracing model synthesizes a multipath \emph{channel impulse response} (CIR). The modeled components comprise: (i) a
line-of-sight (LOS) path when the link is geometrically visible within the corridor, (ii) single-bounce specular wall
reflections, and (iii) a corner-mediated component that captures diffraction around the corner (or a simplified
corner-scatter surrogate). The $m$-th component is assigned a delay $\tau_m$ and a complex coefficient $\alpha_m \propto
a(d_m)e^{-jk_0 d_m}$, $k_0=\frac{2\pi}{\lambda}$ where $d_m$ is the path length and $a(d_m)$ is a distance-dependent
amplitude term (Friis-constant or inverse-distance, depending on configuration). The continuous-time path components are
quantized onto a discrete-time CIR grid with sampling period $T_s=1/B$ and truncated to $L_{\mathrm{tap}}$ taps,
yielding a discrete CIR $h[n]$ for each link.

The reference channel is the direct TX$\rightarrow$UE CIR, denoted $h_0(t,n)$, with clean wideband power
$P_0^{\mathrm{clean}}(t) = \sum_{n} |h_0(t,n)|^2$. For each BD $k$, a cascaded BD channel is formed by convolving
the TX$\rightarrow$BD CIR with the position-dependent BD$\rightarrow$UE CIR and scaling by a BD reflection factor
$\Gamma_k$: $h_k^{\mathrm{casc}}(t,n) = \Gamma_k \Big(h_{\mathrm{tx}\rightarrow k} *
h_{k\rightarrow\mathrm{ue}}(t)\Big)[n]$, $P_k^{\mathrm{clean}}(t) = \sum_{n}
\big|h_k^{\mathrm{casc}}(t,n)\big|^2$.
\noindent Finally, complex baseband observations are generated by adding circular complex Gaussian noise,
$Y_0(t)=\mu_0(t)+W_0(t)$, $Y_k(t)=\mu_k(t)+W_k(t)$, where $W \sim \mathcal{CN}(0,\sigma^2)$ and $\mu(t) \propto
\sqrt{P_{\mathrm{tx}}}\,\sqrt{P^{\mathrm{clean}}(t)}$. The resulting  power measurements are $ Z_0(t)=|Y_0(t)|^2$,
$Z_k(t)=|Y_k(t)|^2$, from which the power observations used by the tracker are formed.

In the simulations, we consider U-shaped and S-shaped trajectories.
The carrier frequency is set to $1.8$~GHz and the system bandwidth to $20$~MHz. After Doppler
separation, the per-tone processing bandwidth is set to $B_t=125$~Hz, corresponding to a matched-filter integration
window of $T_{\mathrm{win}}=8$~ms, consistent with the tracker input rate. The thermal-noise power spectral density is
$N_0=-174.3$~dBm/Hz, the receiver noise figure (NF) is $\mathrm{NF}=6$~dB, and the transmitter power is
$P_{\mathrm{tx}}=23$~dBm. The BD reflection loss is set to $12$~dB and all antenna gains are assumed to be $0$~dBi. In
the tracking code, the log-distance singularity offset is fixed to $d_0=0.1$~m, the candidate exponent set is
$\mathcal{P}=\{1.2,1.3,\ldots,4.0\}$, the motion Huber threshold is set to $\delta_{\mathrm{move}}=\Delta$, the maximum
number of outer alternations is $N_{\mathrm{out}}=5$, and each per-BD offset update uses $N_{\mathrm{IRLS}}=2$ inner
IRLS steps.

\subsection{Algorithm Parameter Sensitivity}

We evaluated the sensitivity of the proposed algorithm to the Huber threshold $\delta\in\{3,4,5\}$~dB,
motion-regularization weight $\lambda_{\mathrm{move}}\in\{3,5,7\}$, and admissible move radius
$R_{\mathrm{move}}\in\{0.4,0.5,0.6\}$~m, using the U-shaped trajectory in all three study areas for grid resolutions of
$0.2$~m and $0.4$~m.

For the $0.2$~m grid, the pooled median error over all cases ranges only from $0.253$~m to $0.272$~m, indicating low
sensitivity to parameter choice. The best setting is $\delta=4$~dB, $\lambda_{\mathrm{move}}=3$, and
$R_{\mathrm{move}}=0.4$~m, with several nearby settings performing nearly identically. For the $0.4$~m grid, the pooled
median error ranges from $0.254$~m to $0.288$~m, and the same parameter set remains optimal. Overall, the results
indicate good robustness to practical parameter variations, with the nominal Huber threshold and a mild motion penalty
giving the best performance.

\subsection{Tracking Performance}

In this section, the proposed algorithm uses the optimal parameter set derived from the sensitivity
analysis. We benchmark the method against a method that uses a cell-averaged fingerprint from the underlying ray-tracing
path-loss map instead of $\Delta_{\ell,k}$  in~\eqref{eq:log-distance-fingerprint} but otherwise uses the same
computational steps as the proposed algorithm. This way, we can isolate the loss due to the log-distance approximation,
apart from residual grid discretization and within-cell averaging. This benchmark allows us to assess the expected
improvement by utilizing better channel models.

\begin{figure}[t]
    \centering
    \setlength{\tabcolsep}{0pt}

    \begin{tabular}{cc}
        \subfloat[]{%
            \includegraphics[width=0.49\columnwidth]{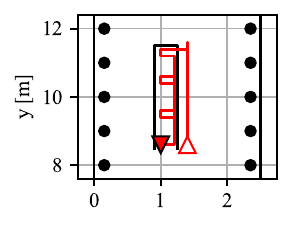}%
            \vspace{-15pt}%
            \label{fig:case2_dx02_prof_u}%
        }
        \addtocounter{subfigure}{-1}
        &
        \subfloat[]{%
            \includegraphics[width=0.47\columnwidth]{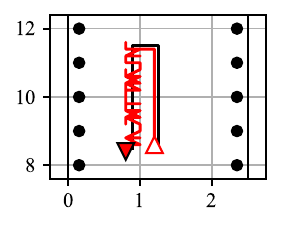}%
            \vspace{-15pt}%
            \label{fig:case2_dx02_oracle_u}%
        }
        \addtocounter{subfigure}{-1}
        \\[-1em]
        \subfloat[]{%
            \includegraphics[width=0.49\columnwidth]{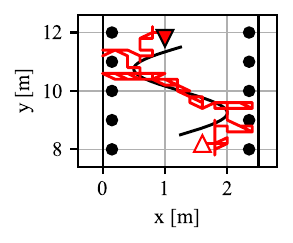}%
            \vspace{-15pt}%
            \label{fig:case2_dx02_prof_s}%
        }
        &
        \subfloat[]{%
            \includegraphics[width=0.47\columnwidth]{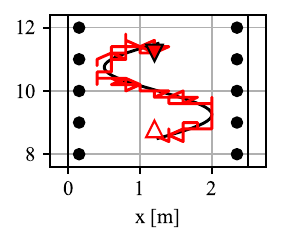}%
            \vspace{-15pt}%
            \label{fig:case2_dx02_oracle_s}%
        }
    \end{tabular}
    \vspace{-2pt}
    \caption{Trajectory-estimation examples for Case~2 with $\Delta = 0.2$~m.
    The top row shows the U-shaped trajectory and the bottom
    row the S-shaped trajectory. In column (a),
    the proposed method, and in (b) the benchmark based on
    ray-tracing-based BD--UE channels.
    Legend: {\color{black}\rule[0.5ex]{1.1em}{1pt}}~true~trajectory;
    {\color{red}\rule[0.5ex]{1.1em}{1pt}}~estimated~trajectory; $\triangle$~start~point;
    $\triangledown$~end~point; $\bullet$~BD.}
    \label{fig:trajectory_overlay_nominal}
\end{figure}

Since Case~2 has more uniform channel conditions compared to the other cases, estimated trajectories in Case~2 for
different grid sizes are shown in Fig.~\ref{fig:trajectory_overlay_nominal}. As can be observed, 
the proposed method can follow the true trajectory similar to the benchmark, although its performance 
is affected by the model mismatch. 

\begin{figure}[t]
    \vspace{-5pt}
    \centering
    \setlength{\tabcolsep}{0pt}

    \begin{tabular}{ccc}
        \subfloat[]{%
            \includegraphics[width=0.36\columnwidth]{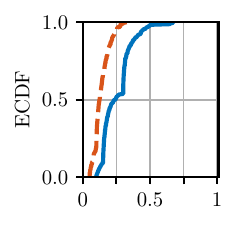}%
            \vspace{-10pt}%
            \label{fig:ecdf_case1_dx02}%
        }
        \addtocounter{subfigure}{-1}
        &
        \subfloat[]{%
            \includegraphics[width=0.31\columnwidth]{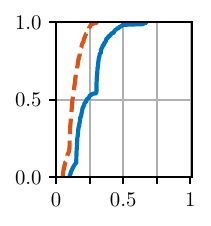}%
            \vspace{-10pt}%
            \label{fig:ecdf_case2_dx02}%
        }
        \addtocounter{subfigure}{-1}
        &
        \subfloat[]{%
            \includegraphics[width=0.31\columnwidth]{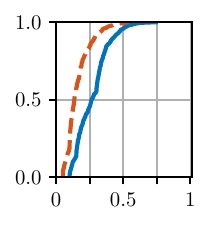}%
            \vspace{-10pt}%
            \label{fig:ecdf_case3_dx02}%
        }
        \addtocounter{subfigure}{-1}
        \\[-1em]
        \subfloat[]{%
            \includegraphics[width=0.36\columnwidth]{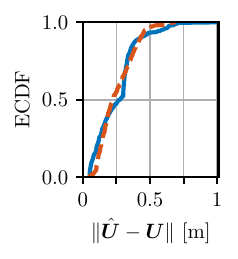}%
            \vspace{-10pt}%
        }
        &
        \subfloat[]{%
            \includegraphics[width=0.32\columnwidth]{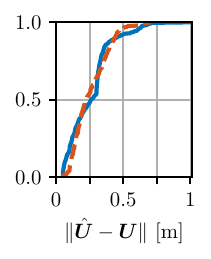}%
            \vspace{-10pt}%
        }
        &
        \subfloat[]{%
            \includegraphics[width=0.32\columnwidth]{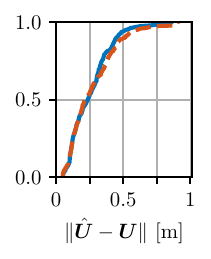}%
            \vspace{-10pt}%
        }
    \end{tabular}
    \caption{Empirical cumulative distribution functions (ECDF) of the positioning error $\|\hat{\bm{U}} - \bm{U}\|_2$
    for the U-shaped trajectory. 
    The top row corresponds to $\Delta = 0.2$ m and the bottom row to $\Delta = 0.4$ m; the columns (a)--(c)
    correspond to Cases 1--3. 
    Legend:
    {\color[rgb]{0.00,0.45,0.74}\rule[0.5ex]{1.1em}{1pt}}~Proposed;
    {\color[rgb]{0.85,0.33,0.10}\rule[0.5ex]{0.22em}{1pt}\hspace{0.12em}\rule[0.5ex]{0.22em}{1pt}\hspace{0.12em}\rule[0.5ex]{0.22em}{1pt}}~Benchmark.}
    \label{fig:ecdf}
\end{figure}

The corresponding empirical cumulative distribution function (ECDF) of the Euclidean distance between true location $\bm{U}$ and the estimated location
$\hat{\bm{U}}$ for all three cases are shown in Fig.~\ref{fig:ecdf} for the U-shaped trajectory. With $0.2\times
0.2~\mathrm{m}^2$ cells, the tracker achieves median errors of approximately $0.23$ -- $0.27$~m in the three study
areas, while the $90$th percentile stays below $0.43$~m. For this resolution, the benchmark algorithm provides a clear
and consistent gain in all three cases: the median error decreases from $0.228$--$0.265$~m to $0.123$--$0.136$~m, and
the $90$th percentile from $0.385$--$0.430$~m to $0.214$--$0.288$~m, which indicates that performance is still partly
limited by model mismatch when resolution is high. With $0.4\times 0.4~\mathrm{m}^2$ cells, the median error remains
similar ($0.24$--$0.28$~m), but the upper tail increases slightly to about $0.43$--$0.45$~m. In this case, the advantage
of the benchmark is much smaller. The median error decreases only from $0.238$--$0.279$~m to $0.218$--$0.225$~m, while
the $90$th percentile becomes comparable to that of the proposed tracker. This suggests that, at the coarser grid,
spatial quantization and the discrete motion model dominate, reducing the benefit of a more accurate fingerprint. Hence,
a finer grid is needed to convert improved channel modeling into a clear localization gain.

Although not depicted here, the estimated pathloss exponent remains physically plausible for indoor propagation; it
decreases from $\hat{p}\approx 2.6$ at the finer grid to $\hat{p}\approx 2.1$--$2.3$ at the coarser grid.

The presented ECDFs in Fig.~\ref{fig:ecdf} show that although the studied U-shaped trajectory is exposed to
different channel conditions when the UE is closer to the origin, the proposed method robustly provides below half-meter
localization accuracy with a modest tuning effort.

\section{Measurement Results}
\begin{figure}[t]
\centering
\includegraphics[width=0.99\columnwidth]{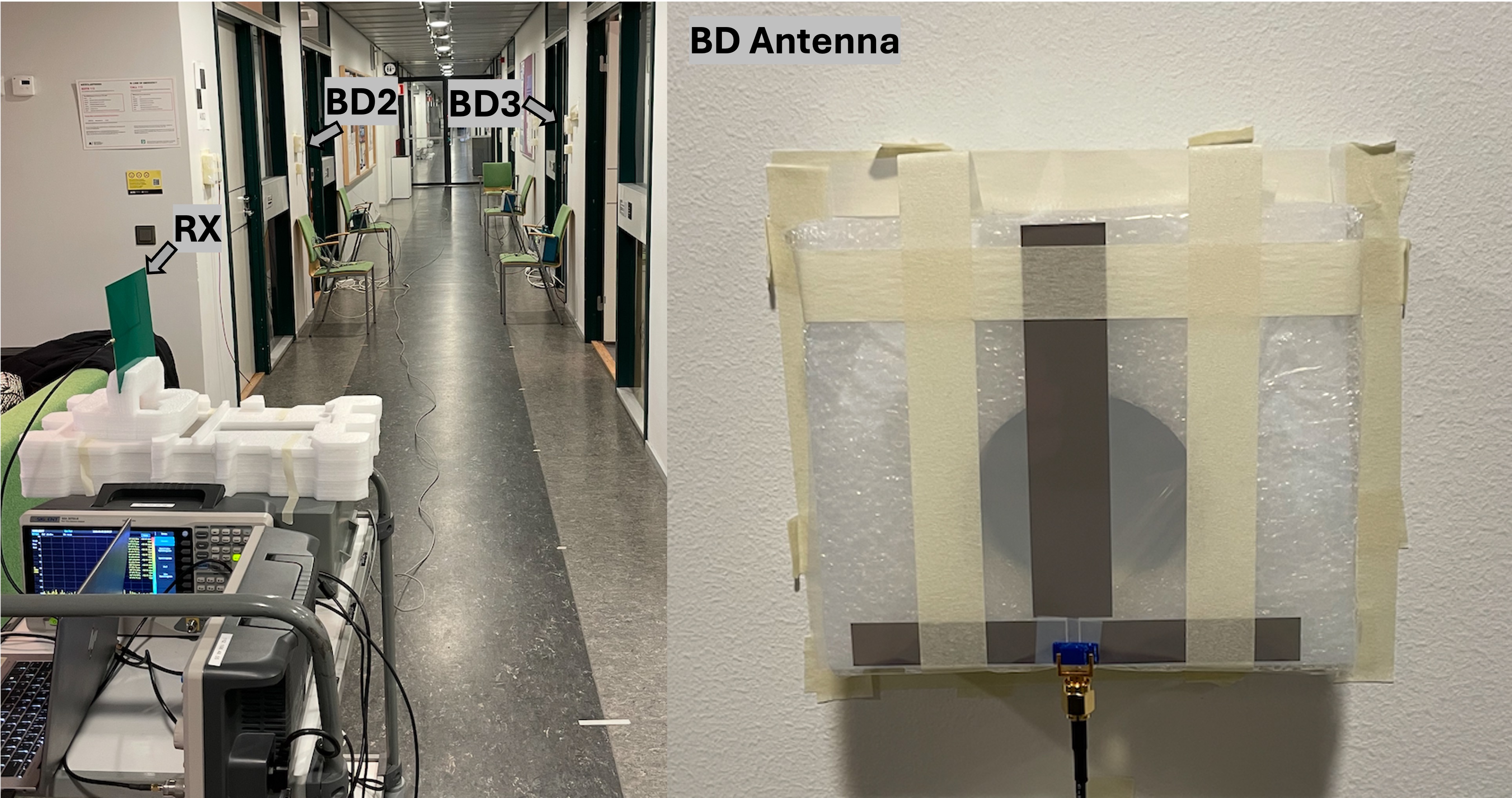}
\caption{Measurement setup showing the RX antenna and the BDs.}
\label{fig:setup}
\end{figure}

Measurements are conducted in an office corridor in the Aalto University Maarintie~8 building under NLOS conditions.
Fig.~\ref{fig:setup} shows the measurement hardware, consisting of the receiver (RX) that represents the UE, four BDs, and a transmitter (TX) that is not visible in Fig.~\ref{fig:setup} because it is placed at a distant NLOS location relative to the RX.  

In a Cartesian coordinate system with $x=0$ on the corridor centerline, the
left wall is at $x=-1$~m, and the right wall at $x=1$~m. The BDs are installed approximately at the height of 
$1.75$~m and their 2D positions are as follows: 
$\mathrm{BD}2:(1,5.0)$, $\mathrm{BD}3:(-1,4.75)$, $\mathrm{BD}4:(-1,7.2)$, and
$\mathrm{BD}5:(1,8.0)$. The TX is located at $(1.5,28.9)$. 

During the experiments, the TX is configured to transmit a continuous-wave sinusoidal signal at $866$~MHz 
with $30$~dBm power. BDs are configured to shift the impinging signal by a specific frequency to separate 
each BD in the frequency domain as follows: 
$\mathrm{BD}2: 6.5\text{ kHz}$, $\mathrm{BD}3: 7.5\text{ kHz}$, $\mathrm{BD}4: 8.5\text{ kHz}$, and
$\mathrm{BD}5: 9.5\text{ kHz}$. The measurements are recorded while the RX is moved to different locations that are
approximately on the centerline (+Y direction) of the defined coordinate system starting from $(0,0)$ up to $(0,14)$~m.
The RX is moved at a constant velocity from the start location to the final point. The RX sampling rate results in
measurements with an approximately constant spatial increment of $0.074$~m.
The same experiment is repeated $92$ times to fully capture the statistical behavior of the proposed setup.

For each experiment, the recorded data are stored in a measurement matrix whose rows correspond to 
successive RX positions along the trajectory, and the columns are different experiments. In order to mitigate 
the reference position uncertainty, the data matrix rows are averaged over five consecutive samples so that the 
spatial granularity is increased to approximately $0.37$~m.    

This experimental campaign calls for the following algorithm parameters: grid cell size to $\Delta=0.4$~m; the 
tracking region $y\in[4,8]$~m, which matches the corridor segment spanned by BD2--BD5; 
$d_0=0.1$~m, $\mathcal{P}=\{1.2,1.3,\ldots,4.0\}$, $\delta=4$~dB, $\lambda_{\mathrm{move}}=5$, $R_{\mathrm{move}}=0.6$~m, $\delta_{\mathrm{move}}=\Delta=0.4$~m, $N_{\mathrm{out}}=5$, and $N_{\mathrm{IRLS}}=2$. The admissible cells are restricted to the rectangular tracking window around 
the corridor centerline so that the cells very close to the walls can be avoided. 

As a low-complexity measurement-domain baseline, we use a weighted-average estimator that computes the UE position as a convex combination
of the BD locations: $\hat{\bm U}(t)=\sum_{k=1}^{K} w_k(t)\,\bm B_k$, with weights proportional to the measured BD
powers: $ w_k(t)={10^{Z_k(t)/10}}({\sum_{k'=1}^{K} 10^{Z_{k'}(t)/10}})^{-1}$, $k=1,\dots,K$. Unlike the simulation study,
the measurement campaign does not provide a calibrated environment model that would support a comparable ray-tracing or
oracle-style benchmark. The measured comparison is therefore restricted to this simple signal-strength baseline.

\begin{figure}[t]
    \centering
    \setlength{\tabcolsep}{0pt}
    \begin{tabular}{cc}
        \subfloat[]{%
            \includegraphics[width=0.49\columnwidth]{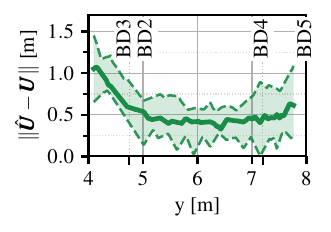}%
            \vspace{-10pt}%
            \label{fig:meas_poserr_866}%
        }
        &
        \subfloat[]{%
            \includegraphics[width=0.51\columnwidth]{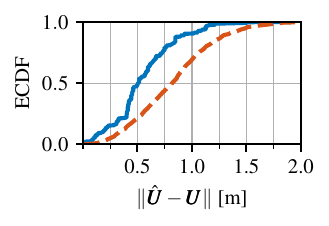}%
            \vspace{-10pt}%
            \label{fig:meas_ecdf_866}%
        }
    \end{tabular}
    \caption{Measurement results: in (a) position error $\|\hat{\bm{U}}-\bm{U}\|_2$ versus reference position on $y$-axis, and in (b) the empirical cumulative distribution function (ECDF) of the positioning error $\|\hat{\bm{U}}-\bm{U}\|_2$ for the proposed algorithm and the baseline.
    Legend:
    (a)
    {\color[rgb]{0.10,0.55,0.28}\rule[0.5ex]{1.1em}{1pt}} Median;
    {\color[rgb]{0.10,0.55,0.28}\rule[0.5ex]{0.22em}{1pt}\hspace{0.12em}\rule[0.5ex]{0.22em}{1pt}\hspace{0.12em}\rule[0.5ex]{0.22em}{1pt}} 10\% and 90\% bounds;
    vertical dotted lines indicate BD $y$-coordinates.
    (b)
    {\color[rgb]{0.00,0.45,0.74}\rule[0.5ex]{1.1em}{1pt}} Proposed;
    {\color[rgb]{0.85,0.33,0.10}\rule[0.5ex]{0.22em}{1pt}\hspace{0.12em}\rule[0.5ex]{0.22em}{1pt}\hspace{0.12em}\rule[0.5ex]{0.22em}{1pt}} Baseline.}
    \label{fig:meas_866_poserr_ecdf}
\end{figure}

Fig.~\ref{fig:meas_866_poserr_ecdf} shows the positioning error results. For each run, the path-loss exponent 
$p$ is selected by a coarse search, which yields experiment-dependent estimates $\hat p\in[2.5,4.0]$ at $866$~MHz.
The estimated per-BD offset parameters remain stable at the few-dB level, consistent with quasi-static illumination
geometry and BD hardware settings. The per-BD gain estimates exhibit standard deviations of
approximately $2.0$--$4.2$~dB across experiments. 

In Fig.~\ref{fig:meas_poserr_866}, the positioning error is shown versus the increasing $y$-coordinate of the RX,
together with the $10\%$ and $90\%$ confidence bounds over $92$ measurements. As can be observed, the error is large
when the RX is at the boundaries and is lowest when all four BDs have significant power. This result shows that the
log-distance aggregate is more accurate when the received BD signals are high enough. Conversely, when the received
power is low, a different model is required. 

Fig.~\ref{fig:meas_ecdf_866} shows the localization-error ECDFs computed over all experiments for the baseline and the
proposed method. The proposed algorithm achieves an aggregated median error of $0.505$~m, with $90$th-percentile error 
of $0.933$~m. The weighted-average baseline remains clearly weaker,
with an aggregated median error of $0.818$~m and corresponding $90$th-percentile error of $1.33$~m. 

Overall, the proposed method provides a clear and repeatable improvement over the
simple weighted-average baseline, and exhibits similar behavior to simulation results shown in
Fig.~\ref{fig:trajectory_overlay_nominal} when the RX has $y$-coordinate in $(5,7.2)$~m. The remaining positioning 
errors are thus primarily attributable to local multipath-induced deviations from the single-slope log-distance 
surrogate and to the limited spatial sampling density in the measurements.

\section{Conclusions}
This paper studied backscatter-assisted indoor NLOS tracking in corridor environments, where passive BDs at known
locations act as virtual anchors through Doppler-separated signatures in standard channel estimates. A
corridor-constrained grid-based MAP/Viterbi tracker was developed using noncoherent power-domain measurements, motion
regularization, and Huber-based estimation of BD-dependent offsets. In the considered corridor scenarios, the method
achieved median errors of about $0.23$--$0.27$~m in simulation and an aggregated median error of $0.505$~m at
$866$~MHz in office-corridor measurements, while outperforming a simple weighted-average baseline. These results show
that passive asynchronous BDs can support practical sub-meter indoor NLOS tracking without BD phase
synchronization or transmit-power calibration. The remaining performance gap is mainly tied to propagation-model
fidelity rather than the tracking architecture itself, which makes the proposed framework a useful basis for richer
environment-aware fingerprints and broader experimental validation in future indoor deployments.

\bibliographystyle{IEEEtran}
\bibliography{references}

\end{document}